\documentclass[nofootinbib,prd,twocolumn,showpacs,showkeys,preprintnumbers]{revtex4-1}
\usepackage{hyperref,amssymb,amsmath,mathrsfs,bm,graphicx}
\usepackage{color}
\begin{document}

\title {Hyperbolically symmetric static fluids: A general study}
\author{L. Herrera}
\email{lherrera@usal.es}
\affiliation{Instituto Universitario de F\'isica
Fundamental y Matem\'aticas, Universidad de Salamanca, Salamanca 37007, Spain}
\author{A. Di Prisco}
\email{alicia.diprisco@ciens.ucv.ve; adiprisc@fisica.ciens.ucv.ve}
\affiliation{Escuela de F\'\i sica, Facultad de Ciencias, Universidad Central de Venezuela, Caracas 1050, Venezuela}
\author{J. Ospino}
\email{j.ospino@usal.es}
\affiliation{Departamento de Matem\'atica Aplicada and Instituto Universitario de F\'isica
Fundamental y Matem\'aticas, Universidad de Salamanca, Salamanca 37007, Spain}
\begin{abstract}

We  carry on a comprehensive study on  static fluid distributions endowed with hyperbolical symmetry. Their physical properties are analyzed in detail. The energy density  appears to be  necessarily negative, which suggests that any possible application of this kind of fluids requires extreme physical conditions where quantum effects are expected to play an important role. Also, it is found that the fluid distribution cannot fill the region close to the center of symmetry.  Such a region may be represented by a vacuum cavity around the center.  A suitable definition of mass function, as well as  the Tolman  mass are explicitly calculated.  While the former is positive defined, the latter is negative in most cases, revealing the repulsive nature of gravitational interaction.
A general approach to obtain exact solutions is presented and some exact analytical solutions are exhibited. 
\end{abstract}
\maketitle

\section{Introduction}
In recent  papers  \cite{1, 2} an alternative  global description of the  Schwarzschild black hole  has been proposed, motivated on the one hand   by   the well known fact that any transformation that maintains the static form of the Schwarzschild metric (in the whole space--time) is unable to remove the coordinate singularity in the line element  \cite{rosen}. 

On the other hand, based on the physically reasonable point of view that  any equilibrium final state of a physical process should  be static, it would be desirable  to have  a static solution over the whole space--time.

However, as  is well known,  no static observers can be defined inside the horizon (see \cite{Rin, Caroll} for a discussion on this point), this conclusion becomes intelligible if we recall  that the Schwarzschild horizon is also a Killing horizon, implying that the time--like Killing vector existing outside the horizon, becomes space--like inside it.

Thus, outside the horizon ($R >2M$) one has the usual Schwarzschild  line element corresponding to the spherically symmetric vacuum solution to the Einstein  equations, which in polar coordinate reads (with signature $+2$)
\begin{eqnarray}
ds^2&=&-\left(1-\frac{2M}{R}\right)dt^2+\frac{dR^2}{\left(1-\frac{2M}{R}\right)}+R^2d\Omega^2, \nonumber \\ d\Omega^2&=&d\theta^2+\sin^2 \theta d\phi^2.
\label{w2}
\end{eqnarray}

This metric is static and spherically symmetric, meaning that it admits four Killing vectors:
\begin{eqnarray}
\mathbf{\chi }_{(\mathbf{0})} = \partial _{\mathbf{t}}, \quad {\bf \chi_{(2)}}=-\cos \phi \partial_{\theta}+\cot\theta \sin\phi \partial_{\phi}\nonumber \\
{\bf \chi_{(1)}}=\partial_{\phi} \quad {\bf \chi_{(3)}}=\sin \phi \partial_{\theta}+\cot\theta \cos\phi \partial_{\phi}.
\label{2cmh}
\end{eqnarray}

However, when $R <2 M$ the signature remains $+2$ but the $g_{tt}$ and the $g_{RR}$ terms switch their signs, which explains the fact that the time--like Killing vector outside the horizon, becomes space--like inside it. Also, an apparent line element singularity  appears at  $R =2 M$. Of course, as is  well known, these drawbacks  can be removed by coordinate transformations, but at the price that, as mentioned before, the staticity is lost within the horizon.

In order to save the staticity inside  the horizon, the   model proposed in \cite{1} describes the space time as consisting of a complete four dimensional manifold  (described by (\ref{w2})) on the exterior side of the horizon and a second (different) complete four dimensional solution in the interior of it.  More specifically, to ensure that the vector $\mathbf{\chi }_{(\mathbf{0})} = \partial _{\mathbf{t}}$ be time--like (inside the horizon),   a change in signature as well as a  change in the symmetry at the horizon was required. The $\theta-\phi$ sub-manifolds have  spherical symmetry on the exterior and hyperbolic symmetry in the interior. The two meet only at  $R=2M$, $\theta=0$.

Thus the model  permits for a change in symmetry, from spherical outside the horizon to hyperbolic inside the horizon. Doing so, one has a static solution everywhere, but the symmetry of the
$R =2 M$ surface is different at both sides of it. We have to stress that we  do not know if there is any specific mechanism behind such a change of symmetry and signature. However,  the main point is that  the change of symmetry (and signature) was the  way followed in \cite{1} to obtain a globally static solution.

Thus, the solution proposed for $R <2 M$ is (with signature $(+ - - -)$):
\begin{eqnarray}
ds^2&=&\left(\frac{2M}{R}-1\right)dt^2-\frac{dR^2}{\left(\frac{2M}{R}-1\right)}-R^2d\Omega^2, \nonumber \\ d\Omega^2&=&d\theta^2+\sinh^2 \theta d\phi^2.
\label{w3}
\end{eqnarray}

This is a static solution with the $(\theta  ,\phi )$ space describing a positive Gaussian curvature.
It admits the four Killing vectors
\begin{eqnarray}
\mathbf{\chi }_{(\mathbf{0})} = \partial _{\mathbf{t}}, \quad {\bf \chi_{(2)}}=-\cos \phi \partial_{\theta}+\coth\theta \sin\phi \partial_{\phi}\nonumber \\
{\bf \chi_{(1)}}=\partial_{\phi} \quad {\bf \chi_{(3)}}=\sin \phi \partial_{\theta}+\coth\theta \cos\phi \partial_{\phi}.
\label{2cmhy}
\end{eqnarray}

A solution to the Einstein equations of the form given by (\ref{w3}), defined by the hyperbolic symmetry  (\ref{2cmhy}), was first considered by Harrison \cite{Ha}, and has been more recently the subject of research  in different contexts (see \cite{ellis, 1n, Ga, Ri, mim, Ka, Ma, ren, mimII} and references therein).

In \cite{2}, a general study of  geodesics in the spacetime described by (\ref{w3}) was presented, leading to some interesting conclusions about the behaviour of  a test   particle in this new picture of the Schwarzschild  black hole. Our purpose in this work is to carry out  a complete study on the physical properties of a fluid distribution in the region inner to the horizon, endowed with the symmetry given by (\ref{2cmhy}). Such a fluid distribution might serve as the source of (\ref{w3}).

The mass function ($m$) as well as the Tolman mass ($m_T$) are defined for our fluid distribution. It is shown that within the region $r<2m$ the energy density is negative, a discussion about the physical implications of this fact is presented. A general approach to obtain any exact solution corresponding to a spacetime admitting hyperbolical symmetry is provided and some examples are found and analyzed.
\section{Basic definitions, notation and equations}
In this section we shall present the physical variables and the relevant equations necessary for describing a static self--gravitating locally anisotropic fluid admitting the four  Killing vectors  (\ref{2cmhy}).

 \subsection{The metric}

We consider hyperbolically  symmetric distributions of  static
fluid, which for the sake of completeness we assume to be locally anisotropic and  which may be (or may be not) bounded from the exterior by a
surface $\Sigma^{e}$ whose equation is $r=r_{\Sigma^{e}}=\rm constant$. On the other hand as we shall see below, the fluid distribution cannot  fill the central region, in which case we may assume that such a region is represented by an empty vacuole, implying that the fluid distribution is also bounded from the inside  by a
surface $\Sigma^{i}$ whose equation is $r=r_{\Sigma^{i}}=\rm constant$.

\noindent
The line element is given in polar  coordinates (with the same signature as (\ref{w3})) by

\begin{equation}
ds^2=e^{\nu} dt^2 - e^{\lambda} dr^2 -
r^2 \left( d\theta^2 + \sinh^2\theta d\phi^2 \right),
\label{metric}
\end{equation}

\noindent
where, due to the imposed symmetry $\nu(r)$ and $\lambda(r)$ are exclusively functions of  $r$. We
number the coordinates: $x^0=t; \, x^1=r; \, x^2=\theta; \, x^3=\phi$.

\noindent
The metric (\ref{metric}) has to satisfy Einstein field equations

\begin{equation}
G^\nu_\mu=8\pi T^\nu_\mu.
\label{Efeq}
\end{equation}
\noindent

Let us next provide a full description of the source.
\subsection{The source}

We shall first consider the most general source, compatible with staticity and  axial symmetry. Afterward we shall impose the hyperbolical symmetry.  

Thus we may write for the
 energy momentum tensor
\begin{eqnarray}
{T}_{\alpha\beta}&=& (\mu+P) V_\alpha V_\beta-P g _{\alpha \beta} +\Pi_{\alpha \beta}.
\label{6bis}
\end{eqnarray}
The above is the canonical algebraic decomposition of a second order symmetric tensor with respect to unit timelike vector, which has the standard physical meaning when $T_{\alpha \beta}$ is the energy-momentum tensor describing some energy distribution, and $V^\mu$ the four-velocity assigned by certain observer.

Then, it is clear that $\mu$ is the energy
density (the eigenvalue of $T_{\alpha\beta}$ for eigenvector $V^\alpha$),  whereas  $P$ is the isotropic pressure, and $\Pi_{\alpha \beta}$ is the anisotropic tensor. We are considering an Eckart frame  where fluid elements are at rest.

Thus, it is immediate to see that

\begin{equation}
\mu = T_{\alpha \beta} V^\alpha V^\beta,
\label{jc10}
\end{equation}

\begin{equation}
P = -\frac{1}{3} h^{\alpha \beta} T_{\alpha \beta},\quad   \Pi_{\alpha \beta} = h_\alpha^\mu h_\beta^\nu \left(T_{\mu\nu} + P h_{\mu\nu}\right),
\label{jc11}
\end{equation}

with $h_{\mu \nu}=g_{\mu\nu}-V_\nu V_\mu$.

The introduction of pressure anisotropy in the study of self-gravitating fluids (Newtonian or relativistic) is justified by the fact that it appears in a large number of physically meaningful situations (see \cite{report} and references therein).
Furthermore, as it has been recently shown \cite{hn}, physical processes of the kind expected in stellar evolution will always tend to produce pressure anisotropy, even if the system is initially assumed to be isotropic. The important point to stress here is that any equilibrium configuration is the final stage of a dynamic  regime and  there is no reason to believe that the acquired anisotropy during this dynamic process,   would disappear in the final equilibrium state, and therefore the resulting configuration, even if  initially  had isotropic pressure, should in principle exhibit pressure anisotropy.

Since we choose the fluid to be comoving in our coordinates, then
\begin{equation}
V^\alpha =(e^{-\nu/2}, 0, 0, 0); \quad  V_\alpha=(e^{\nu/2}, 0, 0, 0).
\label{m1}
\end{equation}

Let us now  define a canonical  orthonormal tetrad (say  $e^{(a)}_\alpha$), by adding to the four--velocity vector $e^{(0)}_\alpha\equiv V_\alpha$, three spacelike unitary vectors

\begin{equation}
e^{(1)}_\alpha\equiv K_{\alpha}=(0, -e^{\lambda/2}, 0, 0); \quad  e^{(2)}_\alpha\equiv L_{\alpha}=\left(0, 0, -r, 0\right),
\label{7}
\end{equation}
\begin{equation}
 e^{(3)}_\alpha\equiv S_{\alpha}=(0, 0, 0, -r\sinh \theta),
\label{3nb}
\end{equation}
with $a=0,\,1,\,2,\,3$ (latin indices labeling different vectors of the tetrad).

The  dual vector tetrad $e_{(a)}^\alpha$  is easily computed from the condition
\begin{equation}
 \eta_{(a)(b)}= g_{\alpha\beta} e_{(a)}^\alpha e_{(b)}^\beta, \qquad e^\alpha_{(a)}e_\alpha^{(b)}=\delta^{(b)}_{(a)},
\end{equation}
where $\eta_{(a)(b)}$ denotes the Minkowski metric.

In order to provide  physical significance to the components of the energy momentum tensor, it is instructive to  apply the Bondi approach \cite{Bo}.

Thus, following Bondi, let us introduce a purely locally
Minkowski frame (l.M.f) with coordinates  ($\tau, x, y, z$) (or equivalently,  consider a tetrad
field attached to such l.M.f.) by:
\begin{eqnarray}
d\tau=e^{\nu/2}dt;\qquad dx=e^{\lambda/2}dr;\nonumber \\ dy=r d\theta;\qquad dz=r\sinh \theta d\phi.
\label{2}
\end{eqnarray}

Denoting by a hat  the components of the energy momentum tensor in such l.M.f.,  we have that the matter content is given by

\begin{equation}
\widehat{T}_{\alpha\beta}= \left(\begin{array}{cccc}\mu    &  0  &   0     &   0    \\0 &  P_{xx}    &   P_{xy}     &   0    \\0       &   P_{yx} & P_{yy}  &   0    \\0       &   0       &   0     &   P_{zz}\end{array} \right) \label{3},
\end{equation}
\\
where $\mu, P_{xy}, P_{xx}, P_{yy}, P_{zz}$ denote the energy density and different stresses, respectively, as measured by our locally defined Minkowskian observer.

This is the general expression for the energy--momentum tensor (in the l.M.f.)  only assuming axial symmetry.  However, as consequence of the  hyperbolical  symmetry of the system, it follows from the Einstein equations that the off diagonal term $P_{xy}$ vanishes and,   in general, $ P_{xx}  \neq  P_{yy}  = P_{zz}$.

The components of our tetrad field in the Minkowski coordinates read
\begin{eqnarray}
&&\hat V_\alpha=(1,0,0,0);\quad \hat K_\alpha=(0,-1,0,0);\quad  \hat L_\alpha=(0,0,-1,0),\nonumber \\
 &&\hat S_{\alpha}=(0, 0, 0, -1),
\label{4}
\end{eqnarray}
from which we may write
\begin{equation}
\widehat{T}_{\alpha\beta}= (\mu+P_{zz})\hat V_\alpha \hat V_\beta-P_{zz} \eta _{\alpha \beta} +(P_{xx}-P_{zz})\hat K_\alpha \hat K_\beta \label{5},
\end{equation}
where $\eta_{\alpha \beta}$ denotes the Minkowski metric.

Then transforming back to the  coordinates of (\ref{metric}), we obtain the components of the energy momentum tensor in terms of the physical variables as defined in the l.M.f.
\begin{equation}
{T}_{\alpha\beta}= (\mu+P_{zz}) V_\alpha V_\beta-P_{zz} g _{\alpha \beta} +(P_{xx}-P_{zz}) K_\alpha  K_\beta.
\label{6}
\end{equation}

It would be useful  to express the anisotropic tensor   in the form

\begin{eqnarray}
\Pi_{\alpha \beta}=\Pi\left(K_\alpha K_\beta+\frac{h_{\alpha
\beta}}{3}\right) \label{6bb},
\end{eqnarray}

with

\begin{eqnarray}
\Pi=P_{xx}-P_{zz},
\label{6bb}
\end{eqnarray}

and
\begin{equation}
P=\frac{P_{xx}+2P_{zz}}{3}.
\label{7P}
\end{equation}
Or, inversely
\begin{eqnarray}
P_{zz}=P-\frac{1}{3}\Pi,
\label{6bc}
\end{eqnarray}
\begin{equation}
P_{xx}=P+\frac{2}{3}\Pi.
\label{6bbcn}
\end{equation}

Since the Lie derivative and the partial derivative commute, then
\begin{equation}
\mathcal{L}_\chi (R_{\alpha \beta}-\frac{1}{2}g_{\alpha \beta}{\cal R})=8\pi \mathcal{L}_\chi T_{\alpha \beta}=0,
\label{ccm1}
\end{equation}
for any $\chi$ defined by  (\ref{2cmhy}), implying that all physical variables only depend on $r$.

If  the fluid is bounded from the exterior by a hypersurface $\Sigma^e$ described by the equation  $r=r_{\Sigma^e}=constant$, then the smooth matching of (\ref{w3}) and (\ref{metric}) on $\Sigma^e$ requires the fulfillment of the Darmois conditions \cite{Darmois}, imposing the continuity of the first and the second fundamental forms, which imply

\begin{equation}
e^{\nu_{\Sigma^e}}=\frac{2M}{r_{\Sigma^e}}-1, \qquad e^{\lambda_{\Sigma^e}}=\frac{1}{\frac{2M}{r_{\Sigma^e}}-1},\qquad P_{xx}(r_{\Sigma^e})=0,
\label{j1}
\end{equation}
and the continuity of the mass function $m(r)$ defined below. 
If we assume that the central region is surrounded by an empty cavity whose delimiting surface is $r=r_{\Sigma^i}=constant$, then the fulfillment of Darmois conditions on $\Sigma^i$ implies
\begin{equation}
e^{\nu_{\Sigma^i}}=1, \qquad e^{\lambda_{\Sigma^i}}=1,\qquad P_{xx}(r_{\Sigma^i})=0,
\label{j2}
\end{equation}
and $m(r_{\Sigma^i})=0$.

If either of conditions above (or both) are not satisfied, then we have to resort to Israel conditions \cite{17}, implying that thin shells are present at either boundary surface (or both).

\subsection{The Einstein equations}
\noindent The non--vanishing components of the Einstein equations for the metric  (\ref{metric}) and the energy momentum tensor  (\ref{6}) are

\begin{eqnarray}
  8\pi \mu &=& -\frac{(e^{-\lambda}+1)}{r^2}+\frac{\lambda ^\prime}{r}e^{-\lambda},\label{mu} \\
  8\pi P_r &=&\frac{(e^{-\lambda}+1)}{r^2}+\frac{\nu ^\prime}{r}e^{-\lambda}, \label{pr}\\
  8\pi P_\bot&=& \frac{e^{-\lambda}}{2}\left (\nu^{\prime\prime}+\frac{{\nu^\prime}^2}{2}-\frac{\lambda^\prime \nu^\prime}{2}+\frac{\nu^\prime}{r}-\frac{\lambda^\prime}{r}\right),\label{pbot}
\end{eqnarray}
where we have used the standard notation $P_{xx}\equiv P_r$ and $P_{zz}=P_{yy}\equiv P_\bot$, and primes denote derivatives with respect to $r$.

It is worth stressing the differences between these equations and the corresponding to the spherically symmetric case (see for example eqs.(2--4) in \cite{16}).

From the equations above or using the conservation laws $T^\alpha_{\beta;\alpha}=0$ we obtain, besides the identity $\dot\mu = 0$ (where dot denotes derivative with respect to $t$), the corresponding hydrostatic  equilibrium equation (the generalized Tolman--Oppenheimer--Volkoff equation)
\begin{eqnarray}
  P_r^\prime+(\mu+P_r)\frac{\nu^\prime}{2}+\frac{2}{r} \Pi&=& 0\label{CPx}.
\end{eqnarray}

Let us now define the mass function $m=m(r)$.
For doing so, let us notice that using (\ref{w3}) we have that  outside the fluid distribution (but inside the horizon)
\begin{equation}
M=-\left(\frac{R}{2} \right)R^3_{232},
\label{m1n}
\end{equation}
where the Riemann tensor component $R^3_{232}$, has been calculated with (\ref{w3}).

Then generalizing the above definition of mass  for the interior of the fluid distribution we may write
\begin{equation}
m(r)=-\left(\frac{r }{2}\right)R^3_{232}=\frac{r (1+e^{-\lambda})}{2}
\label{m3n}
\end{equation}
where now the Riemann tensor component is calculated with (\ref{metric}).

Feeding back (\ref{m3n}) into (\ref{mu}) we obtain

\begin{equation}
m^\prime(r)=-4\pi r^2 \mu \Rightarrow m=-4\pi \int^r_0{\mu r^2dr}.
\label{m2}
\end{equation}

Since $m$ as defined by (\ref{m3n}) is a positive quantity,  then $\mu$ should be  negative  and therefore the weak energy condition is violated, a result already obtained in \cite{mimII}. However it is important to stress that our definition of mass function differs from the one introduced in \cite{mimII}. In particular our $m$ is positive defined whereas the expression used in \cite{mimII} is negative (for the hyperbolically symmetric fluid). 

 The following   comments are  in order at this point.  
 \begin{itemize}
 \item It is worth noticing that while the total energy (mass) of the bounded distribution is unique $(M)$, the definition of the energy localized in a given piece of the fluid distribution is not. As a matter of fact, this ambiguity in the localization of energy, which  is present even in classical electrodynamics \cite{fey}, has been extensively discussed in general relativity, leading to different definitions of energy (see for example \cite{cv, 11, cs, haw, pen, lv, by} and references therein).
 
 \item Our main motivation to study hyperbolically symmetric fluids is directly related to the black hole picture described in the Introduction, according to which the region interior to the horizon is described by (\ref{w3}), whereas the spacetime outside the horizon is described by the usual Schwarzschild metric (\ref{w2}). Now, the parameter $M$ appearing in (\ref{w3}) is the same as the $M$ appearing in  (\ref{w2}), i.e. the total mass of the source, which is positive defined. Then, if we wish that our mass function be continuous at the boundary surface of our fluid distribution, which in turn is the source of (\ref{w3}), it is natural to assume (\ref{m3n}) as the definition  of the mass function. However it should be clear that any other scenario different from the one described above, allows for  other alternative definitions of mass function.

\item The equation (\ref{m3n}), as expected,  is at variance with the definition in the spherically symmetric case ($e^{-\lambda}=1-\frac{2m}{r}$), since we are interested in the region $2m>r$. 

\item If the energy density is regular everywhere then  the mass function must vanish at the center as $m\sim r^3$,  this implies (as it follows from (\ref{m3n})) that the fluid cannot fill the space in the neighborhood of  the center, i.e. there is a cavity around the center which may be, either  empty,  or  filled with a fluid distribution non endowed with hyperbolical symmetry. Thus the hyperbolically symmetric fluid spans from a minimal value of the coordinate $r$ until its external boundary. For the extreme case $\mu=constant$, this minimal value $r_{min.}$ is defined by $-\frac{8\pi}{3} \mu r^2_{min.}>1$. Obviously, if the  energy density is singular in the neighborhood of the center, then this region must also be excluded by physical reasons.
\end{itemize}

From the  above it follows that,  strictly speaking,  we should write instead of (\ref{m2})

\begin{equation}
 m=4\pi \int^r_{r_{min}}{\vert \mu \vert r^2dr},
\label{m3}
\end{equation}
where due to the fact that $\mu$ is negative, we have replaced it by $-\vert \mu \vert$ (as we shall do from now on).

The situation described above is fully consistent with the results obtained in \cite{2} where it was shown that test particles cannot reach the center for any finite value of its energy.

Next, using (\ref{pr}) and (\ref{m3n}) we obtain
\begin{equation}
\nu^\prime=2\frac{4\pi r^3 P_r-m}{r(2 m-r)},
\label{m3}
\end{equation}
from which we may write (\ref{CPx}) as

\begin{eqnarray}
  P_r^\prime+(P_r-\vert \mu\vert)\frac{4\pi r^3 P_r-m}{r(2 m-r)}+\frac{2}{r} \Pi&=& 0\label{m4}.
\end{eqnarray}

This is the hydrostatic equilibrium equation for our fluid. Let us analyze in some detail  the physical meaning of its different terms. The first term is just the gradient of pressure, which is usually negative  and opposing gravity. The second term describes the gravitational ``force'' and contains two different contributions: on the one hand the term $P_r-\vert \mu \vert$ which we expect to be negative (or zero for the stiff equation of state) and is usually interpreted as the ``passive gravitational mass density'' (p.g.m.d.), and on the other hand  the term $4\pi r^3 P_r-m$  that is proportional to the ``active gravitational mass''  (a.g.m.), and which  is negative if $4\pi r^3 P_r<m$. Finally the third term describes the effect of the pressure anisotropy, whose sign depends on the difference between principal stresses.
Two important remarks are in order at this point:
\begin{itemize}
\item It is worth stressing that while the self--regenerative pressure effect (described by the  $4\pi r^3 P_r$ term in (\ref{m4})) has the same sign  as in the spherically symmetric case, the mass function contribution in the second term has the opposite sign with respect to the latter case. This of course is due to the fact that the energy density is negative.
\item If, both, the p.g.m.d. and the a.g.m. are negative, the final effect of the gravitational interaction would be as usual, to oppose the negative pressure gradient. However, because of the equivalence principle, a negative  p.g.m.d. implies a negative inertial mass, which in turn implies that the hydrostatic force term (the pressure gradient and the anisotropic term), and the gravitational force term, switch their roles with respect to the positive energy density  case.
\end{itemize}
\subsection{The Riemann and the Weyl tensor}
 As is well known, the
Riemann tensor may be expressed through the Weyl tensor
$C^{\rho}_{\alpha
\beta
\mu}$, the  Ricci tensor $R_{\alpha\beta}$ and the scalar curvature ${\cal R}$,
as
$$
R^{\rho}_{\alpha \beta \mu}=C^\rho_{\alpha \beta \mu}+ \frac{1}{2}
R^\rho_{\beta}g_{\alpha \mu}-\frac{1}{2}R_{\alpha \beta}\delta
^\rho_{\mu}+\frac{1}{2}R_{\alpha \mu}\delta^\rho_\beta$$
\begin{equation}
-\frac{1}{2}R^\rho_\mu g_{\alpha
\beta}-\frac{1}{6}{\cal R}(\delta^\rho_\beta g_{\alpha \mu}-g_{\alpha
\beta}\delta^\rho_\mu).
\label{34}
\end{equation}

In   our case, the magnetic part of the Weyl tensor vanishes and we can express the Weyl tensor in terms of its electric part ($E_{\alpha \beta}=C_{\alpha \gamma \beta
\delta}V^{\gamma}V^{\delta}$) as
\begin{equation}
C_{\mu \nu \kappa \lambda}=(g_{\mu\nu \alpha \beta}g_{\kappa \lambda \gamma
\delta}-\eta_{\mu\nu \alpha \beta}\eta_{\kappa \lambda \gamma
\delta})V^\alpha V^\gamma E^{\beta \delta},
\label{40}
\end{equation}
with $g_{\mu\nu \alpha \beta}=g_{\mu \alpha}g_{\nu \beta}-g_{\mu
\beta}g_{\nu \alpha}$,   and $\eta_{\mu\nu \alpha \beta}$ denoting the Levi--Civita tensor.

\noindent  The electric part of the Weyl tensor for our metric (\ref{metric}) may be written as
\begin{equation}
  E_{\alpha\beta}=\mathcal{E}\left(K_\alpha K_\beta+\frac{1}{3}h_{\alpha\beta}\right), \label{e1}
\end{equation}
satisfying the following properties:
 \begin{eqnarray}
 E^\alpha_{\,\,\alpha}=0,\quad E_{\alpha\gamma}=
 E_{(\alpha\gamma)},\quad E_{\alpha\gamma}V^\gamma=0,
  \label{propE}
 \end{eqnarray}
\noindent where
\begin{equation}\label{PEE}
  \mathcal{E}=-\frac{e^{-\lambda}}{4}\left(\nu^{\prime\prime}+\frac{ {\nu^\prime}^2}{2}-\frac{\lambda^\prime\nu^\prime}{2}-\frac{\nu^\prime}{r}+\frac{\lambda ^\prime}{r}\right)-\frac{1}{2r^2}\left(1+e^{-\lambda}\right).
\end{equation}

Using the field equations (\ref{mu})--(\ref{pbot}), (\ref{m3n}) and (\ref{PEE}) the following relationship may be obtained
\begin{equation}\label{mE}
  \frac{3m}{r^3}=4\pi\vert \mu \vert +4\pi \Pi-\mathcal{E}.
\end{equation}

\noindent Taking the $r$-derivative of the expression above and using (\ref{m2}) we find

\begin{equation}
{\cal E}= \frac{4\pi}{r^3} \int^r_0{\tilde r^3 \vert \mu \vert ' d\tilde r} +
4\pi \Pi.
\label{Wint}
\end{equation}

\noindent
Finally, inserting (\ref{Wint}) into (\ref{mE}) we obtain

\begin{equation}
m(r) = \frac{4\pi}{3} r^3 \vert \mu \vert -
\frac{4\pi}{3} \int^r_0{\tilde r^3 \vert \mu \vert  'd\tilde r}.
\label{mT00}
\end{equation}

Equation  (\ref{Wint}) relates the Weyl tensor to two fundamental physical properties of the fluid distribution, namely: energy density inhomogeneity and local anisotropy of pressure, whereas  (\ref{mT00}) expresses the mass function in  terms of its value in the case of a homogeneous energy density distribution, plus the change induced by energy density inhomogeneity. In the expressions (\ref{Wint}), (\ref{mT00}) above it must be kept in mind that the center of the distribution should be excluded.

\subsection{Tolman mass}
An alternative definition to  describe the energy content  of a  fluid sphere  was proposed by Tolman many years ago \cite{11} .
\noindent
The Tolman mass generalized for any fluid element of  our  static fluid distribution  inside $\Sigma^e$ reads

\begin{eqnarray}\label{masaT}
  m_T&=&\int _{0}^{2\pi} \int_{0}^\pi \int_{0}^{r}\sqrt{-g}(T^0_0-T^1_1-2T^2_2)d\tilde r d\theta d\phi \\\nonumber
  &=&2\pi (cosh\pi-1)\int_{0}^{r}e^{(\nu+\lambda)/2}\tilde{r}^2(-\vert \mu \vert +P_{r}+2P_{\bot})d\tilde{r}.
\end{eqnarray}

\noindent Using the field equations (\ref{mu})--(\ref{pbot}), the integration of (\ref{masaT})  produces
\begin{equation}\label{masaT1}
  m_T=\frac{(cosh\pi-1)}{4}e^{(\nu-\lambda)/2}r^2\nu^\prime,
\end{equation}

or combining   (\ref{m3}) with  (\ref{masaT1})

\begin{equation}\label{masaT2}
 m_T= \frac{(cosh\pi-1)}{2}e^{(\nu+\lambda)/2}(4\pi P_{r} r^3-m).
\end{equation}

In the light of equation (\ref{masaT2}) (or (\ref{masaT1})) and (\ref{m4})(or (\ref{CPx})), the usual physical interpretation of $m_T$ as a measure of the active gravitational mass becomes evident. It must be stressed the fact that this quantity is negative provided  $4\pi P_{r} r^3<m$, which would imply the  repulsive character of the gravitational interaction in the spacetime under consideration. 

Indeed, let us consider the  four--acceleration  $a_\alpha$, defined as usually by 

\begin{equation}\label{ac}
  a_\alpha=V_{\alpha;\beta}V^\beta,
\end{equation}
that in our case may be written as 
\begin{equation}\label{acb}
  a_\alpha=aK_\alpha,
\end{equation}
where
\begin{equation}\label{aes}
a=\frac{\nu^\prime}{2}e^{-\lambda/2},
\end{equation}

which allows to  write
\begin{equation}\label{aes1}
a=\frac{2m_T}{r^2}\frac{e^{-\nu/2}}{(cosh\pi-1)}.
\end{equation}
\\
Thus  the four-acceleration is directed inwardly (if $4\pi P_{r} r^3<m$). Now, let us recall that $a^\mu$ represents the inertial radial acceleration which is necessary in order to maintain static the frame by canceling the gravitational acceleration exerted on the frame.  Therefore  the fact that  the four--acceleration  is directed radially inward, reveals  the repulsive nature of the gravitational force.

\noindent Next,  taking the $r$-derivative of (\ref{masaT}) and using  (\ref{masaT2}) we find
\begin{widetext}
\begin{equation}\label{masaT3}
  m_T^\prime-\frac{3}{r}m_T=-\frac{(cosh\pi-1)}{2} e^{(\nu+\lambda)/2}r^2\left (4\pi \Pi+{\cal E}\right),
\end{equation}
whose integration produces

\begin{equation}\label{masaT6}
m_T=(m_T)_{\Sigma^e}\left(\frac{r}{r_{\Sigma^e}}\right)^3+\frac{(cosh\pi-1)}{2} r^3\int_{r}^{r_{\Sigma^e}} \frac{e^{(\nu+\lambda)/2}}{\tilde{r}}\left (\mathcal{E}+4\pi \Pi\right) d\tilde{r},
\end{equation}

or using (\ref{Wint})
\begin{equation}\label{masaT5}
m_T=(m_T)_{\Sigma^e}\left(\frac{r}{r_{\Sigma^e}}\right)^3+\frac{(cosh\pi-1)}{2} r^3\int_{r}^{r_{\Sigma^e}} \frac{e^{(\nu+\lambda)/2}}{\tilde{r}}\left (\frac{4\pi}{\tilde{r}^3}\int_{0}^{\tilde{r}}\vert \mu \vert^\prime s^3ds+8\pi \Pi\right )d\tilde{r}.
\end{equation}
\end{widetext}
The above expressions are equivalent to the ones obtained for the spherically symmetric case \cite{14}.

We shall next present the orthogonal splitting of the Riemann tensor, and express it in terms of the variables considered so far. Doing so we shall be able to define the structure scalars for our fluid distribution.

\section{THE ORTHOGONAL SPLITTING OF THE RIEMANN TENSOR AND THE STRUCTURE SCALARS}
Following the orthogonal splitting scheme of the Riemann tensor  considered by Bel \cite{18},
let us introduce the following tensors  (we shall follow closely, with some changes, the notation  of  \cite{19}),
\begin{equation}
Y_{\alpha \beta}=R_{\alpha \gamma \beta \delta}V^{\gamma}V^{\delta},
\label{electric}
\end{equation}
\begin{equation}
Z_{\alpha \beta}=^{*}R_{\alpha \gamma \beta
\delta}V^{\gamma}V^{\delta}= \frac{1}{2}\eta_{\alpha \gamma
\epsilon \mu} R^{\epsilon \mu}_{\quad \beta \delta} V^{\gamma}
V^{\delta}, \label{magnetic}
\end{equation}
\begin{equation}
X_{\alpha \beta}=^{*}R^{*}_{\alpha \gamma \beta \delta}V^{\gamma}V^{\delta}=
\frac{1}{2}\eta_{\alpha \gamma}^{\quad \epsilon \mu} R^{*}_{\epsilon
\mu \beta \delta} V^{\gamma}
V^{\delta},
\label{magneticbis}
\end{equation}
where $*$ denotes the dual tensor, i.e.
$R^{*}_{\alpha \beta \gamma \delta}=\frac{1}{2}\eta_{\epsilon \mu \gamma \delta}R_{\alpha \beta}^{\quad \epsilon \mu}$.

It can be shown that the Riemann tensor  can be expressed through  these tensors in what is called the orthogonal splitting of the Riemann tensor (see \cite{19} for details).
However, instead of using the explicit form of the splitting of Riemann tensor (eq.(4.6) in \cite{19}), we shall proceed as follows (for details see \cite{20}, where the general non--static case has been considered).

Using the Einstein equations we may write (\ref{34}) as
\begin{equation}
R^{\alpha \gamma}_{\quad \beta\delta}=C^{\alpha\gamma}_{\quad
\beta \delta}+28\pi T^{[\alpha}_{\,\,
[\beta}\delta^{\gamma]}_{\,\, \delta]}+8\pi T(\frac{1}{3} \delta
^{\alpha}_{\,\, [\beta}\delta^{\gamma}_{\,\, \delta]}-\delta
^{[\alpha}_{\quad [\beta}\delta^{\gamma]}_{\,\,
\delta]}),\label{RiemannT}
\end{equation}
then feeding back (\ref{6bis}) into  (\ref{RiemannT}) we split the Riemann tensor as
\begin{equation}
R^{\alpha \gamma}_{\,\, \beta \delta}=R^{\alpha
\gamma}_{(I)\,\,\beta\delta}+R^{\alpha
\gamma}_{(II)\,\,\beta\delta}+R^{\alpha\gamma}_{(III)\,\,\beta\delta},
\label{Riemann}
\end{equation}
where
\begin{eqnarray}
 R^{\alpha \gamma}_{(I)\,\,\beta \delta}=16\pi  \mu
V^{[\alpha}V_{[\beta}\delta^{\gamma]}_{\,\,\delta]}-16\pi
Ph^{[\alpha}_{\,\,[\beta}\delta^{\gamma]}_{\,\, \delta]}\nonumber \\+8\pi
( \mu-3
P)(\frac{1}{3}\delta^{\alpha}_{\,\,[\beta}\delta^{\gamma}_{\,\,\delta]}
-\delta^{[\alpha}_{\,\,[\beta}\delta^{\gamma]}_{\,\,\delta]})
\end{eqnarray}
\begin{eqnarray}
R^{\alpha \gamma}_{(II)\,\,\beta\delta}=16\pi \Pi^{[\alpha}_{\,\,
[\beta}\delta^{\gamma]}_{\,\, \delta]}
\end{eqnarray}
\begin{equation}
R^{\alpha
\gamma}_{(III)\,\,\beta\delta}=4V^{[\alpha}V_{[\beta}E^{\gamma]}_{\,\,\,
\delta]}-\epsilon^{\alpha \gamma}_{\quad
\mu}\epsilon_{\beta\delta\nu}E^{\mu\nu}
\label{RiemannH}
\end{equation}
with
\begin{eqnarray}
\epsilon_{\alpha\gamma\beta}=V^\mu\eta_{\mu\alpha\gamma\beta},\quad
\epsilon_{\alpha\gamma\beta}V^\beta=0 \label{conn},
 \end{eqnarray}
and where the vanishing of the  magnetic part of the Weyl tensor ($H_{\alpha \beta}=^{*}C_{\alpha \gamma \beta
\delta}V^{\gamma}V^{\delta}$)  has been used.

Using the results above, we can now  find the explicit expressions for the three tensors $Y_{\alpha \beta}, Z_{\alpha \beta}$ and  $X_{\alpha \beta}$ in terms of  the physical variables, we obtain

\begin{equation}
Y_{\alpha\beta}=\frac{4\pi}{3}(\mu+3
P)h_{\alpha\beta}+4\pi \Pi_{\alpha\beta}+E_{\alpha\beta},\label{Y}
\end{equation}

\begin{equation}
Z_{\alpha\beta}=0,\label{Z}
\end{equation}

and

\begin{equation}
X_{\alpha\beta}=-\frac{8\pi}{3} \vert \mu \vert 
h_{\alpha\beta}+4\pi
 \Pi_{\alpha\beta}-E_{\alpha\beta}.\label{X}
\end{equation}

As shown in \cite{20}, the tensors  above may be expressed in terms of some scalar functions, referred to as structure scalars, by decomposing them into their trace--free part and their trace, as
\begin{equation}
X_{\alpha \beta}=X_T \frac{h_{\alpha \beta}}{3} +X_{TF}\left(K_\alpha K_\beta +\frac{h_{\alpha \beta}}{3}\right),\label{sc1}
\end{equation}

\begin{equation}
Y_{\alpha \beta}=Y_T \frac{h_{\alpha \beta}}{3} +Y_{TF}\left(K_\alpha K_\beta +\frac{h_{\alpha \beta}}{3}\right).\label{sc2}
\end{equation}
These scalars in turn may written as:
\begin{equation}
 X_T=-8\pi  \vert \mu \vert ,
\label{esnIII}
\end{equation}

\begin{equation}
X_{TF}= 4\pi \Pi-{\cal E},
\label{defXTF}
\end{equation}
or using (\ref{Wint})
\begin{equation}
X_{TF}= -\frac{4\pi}{r^3} \int^r_0{\tilde r^3 \vert \mu \vert' d\tilde r},
\label{defXTFbis}
\end{equation}

and
\begin{equation}
Y_T=4\pi( -\vert \mu \vert +3 P),
\label{esnV}
\end{equation}

\begin{equation}
Y_{TF}= 4\pi \Pi+{\cal E},
\label{defYTF}
\end{equation}
or using (\ref{Wint})
\begin{equation}
Y_{TF}=8\pi \Pi+\frac{4\pi}{r^3} \int^r_0{\tilde r^3 \vert \mu \vert' d\tilde r}.
\label{defYTFbis}
\end{equation}

From the above it follows that local anisotropy of pressure is  determined by  $X_{TF}$ and $Y_{TF}$
by
\begin{equation}
8\pi \Pi=X_{TF} + Y_{TF}.
\label{defanisxy}
\end{equation}

To establish the physical meaning of  $Y_T$  and $Y_{TF}$ let us get back to equations (\ref{masaT}) and  (\ref{masaT5}), using (\ref{esnV}) and (\ref{defYTFbis}) we get

\begin{eqnarray}
m_T  &=&  (m_T)_{\Sigma^e} \left(\frac{r}{r_{\Sigma^e}}\right)^3\nonumber \\
 &+& \frac{(cosh\pi-1)}{2} r^3 \int^{r_{\Sigma^e}}_r{\frac{e^{(\nu+\lambda)/2}}{\tilde r} Y_{TF}d\tilde r},
\label{emtebis}
\end{eqnarray}
and

\begin{eqnarray}
m_T = & &  \frac{(cosh\pi-1)}{2} \int^{r}_{0}{\tilde r^2 e^{(\nu+\lambda)/2}
Y_Td\tilde r}.
\label{TolinIII}
\end{eqnarray}

We see that $Y_{TF}$ encompasses the influence of the local anisotropy of pressure and density inhomogeneity on the Tolman mass. Or, in other words, $Y_{TF}$ describes how these two factors modify the value of the Tolman mass, with respect to its value for the homogeneous isotropic fluid. This fact  was at the origin of the definition of complexity provided in \cite{com}.  Indeed, if we assume that the homogeneous (in the energy density) fluid with isotropic pressure is endowed with minimal complexity, then the variable responsible for measuring complexity, which we call the complexity factor, should vanish for this kind of fluid distribution, as it happens for $Y_{TF}$.

Also, it is worth noticing that from (\ref{TolinIII})  $Y_T$ appears to be proportional to the Tolman mass ``density''.

\section{All Static Solutions}
We shall next present a general formalism to express any static hyperbolically symmetric solution in terms of two generating functions. Afterward we shall also present some explicit solutions and their generating functions. The procedure is similar to the one proposed for the spherically symmetric case (see \cite{16}, \cite{16b}).

\noindent Thus, from (\ref{pr}) and  (\ref{pbot}) we may write
\begin{eqnarray}
   8\pi (P_r-P_\bot) &=&\frac{(e^{-\lambda}+1)}{r^2} \\ \nonumber
   &-&\frac{e^{-\lambda}}{2}\left (\nu^{\prime\prime}+\frac{{\nu^\prime}^2}{2}-\frac{\lambda^\prime \nu^\prime}{2}-\frac{\nu^\prime}{r}-\frac{\lambda^\prime}{r}\right),\label{1all}
\end{eqnarray}

\noindent which, by  introducing the following auxiliary functions
\begin{equation}
  \frac{\nu^\prime}{2}=z-\frac{1}{r}\quad;\quad e^{-\lambda}=y,
\end{equation}
\noindent becomes

\begin{equation}
y^\prime+y\left(\frac{2z^\prime}{z}+2z-\frac{6}{r}+\frac{4}{zr^2}\right)=\frac{2}{z}\left(\frac{1}{r^2}-8\pi\Pi\right)\label{dp1}.
\end{equation}

\noindent The integration of (\ref{dp1}) produces
\begin{equation}\label{Allambda}
  e^{\lambda(r)}=\frac{z^2 e^{\int (\frac{4}{zr^2}+2z)dr}}{r^6\left (2\int \frac{z(1-8\pi \Pi r^2)}{r^8}e^{\int (\frac{4}{zr^2}+2z)dr}dr+C_1 \right )}.
\end{equation}

Therefore, any static solution is fully described by the two generating functions $\Pi$ and $z$.

\noindent For the physical variables we  may write

\begin{equation}\label{amu}
4\pi \vert \mu \vert =\frac{m^\prime}{r^2},
\end{equation}

\begin{equation}\label{apr}
  4\pi P_r=\frac{z(2mr-r^2)-m+r}{r^3},
\end{equation}

\begin{equation}\label{apbot}
  8\pi P_\bot=\left(\frac{2m}{r}-1\right)\left(z^\prime+z^2-\frac{z}{r}+\frac{1}{r^2}\right)+z\left(\frac{m^\prime}{r}-\frac{m}{r^2}\right).
\end{equation}

We shall now find some explicit solutions and their respective  generating functions.

\subsection{The conformally flat solutions}
Due to the conspicuous role played by the Weyl tensor in the structure of the fluid distribution, as indicated by (\ref{Wint}) and (\ref{masaT6}), it is worth considering the special case ${\cal E}=0$ (conformal flatness).

Thus we  shall now proceed to integrate the condition
\begin{equation}
{\cal E}=0,
\label{W0}
\end{equation}
which, using (\ref{PEE}), may be written as
\begin{equation}
\left(\frac{e^{-\lambda}\nu'}{2r}\right)'
+ e^{-(\nu+\lambda)} \left(\frac{e^{\nu}\nu'}{2r}\right)' -
\left(\frac{e^{-\lambda}+1}{r^2}\right)' = 0.
\label{prn}
\end{equation}

Introducing the new variables
\begin{equation}
y = e^{-\lambda} \qquad ; \qquad \frac{\nu'}{2} = \frac{u'}{u}
\label{cv}
\end{equation}
the equation(\ref{prn}) is cast into
\begin{equation}
y' +
\frac{2y \left(u''- u'/r + u/r^2 \right)}{\left(u' - u/r\right)} +
\frac{2u}{r^2 \left(u' - u/r\right)} = 0,
\label{yp}
\end{equation}
whose formal solution is
\begin{equation}
y = e^{-\int{k(r)dr}} \left[\int{e^{\int{k(r)dr}}f(r)dr}+ C_1\right],
\label{y}
\end{equation}
where $C_1$ is a constant of integration, and
\begin{equation}
k(r) = 2 \frac{d}{dr}\left[\ln{\left(u'- \frac{u}{r}\right)}\right],
\label{g}
\end{equation}
\begin{equation}
f(r) = -\frac{2u}{r^2 \left(u'- u/r\right)}.
\label{f}
\end{equation}
Changing back to the original variables, eq.(\ref{y}) becomes
\begin{equation}
\frac{\nu'}{2} - \frac{1}{r} = \frac{e^{\lambda/2}}{r}
\sqrt{\alpha r^2 e^{-\nu}-1},
\label{1}
\end{equation}
where $\alpha$ is a constant of integration which using the matching conditions (\ref{j1}) becomes
\begin{equation}
\alpha =\frac{M(9 M-4 r_{\Sigma^e})}{r^4_{\Sigma^e}}.
\label{1nue}
\end{equation}

Next, (\ref{1}) may be formally integrated, to obtain
\begin{equation}
e^\nu = \alpha r^2 \sin^2{\left(\int{\frac{e^{\lambda/2}}{r}dr}+ \gamma\right)}
\label{fint}
\end{equation}
where $\gamma$ is a constant of integration which using (\ref{j1}) reads,
\begin{equation}
\gamma =\arcsin\left[r_{\Sigma^e} \sqrt{\frac{\left(\frac{2M}{r_{\Sigma^e}}-1 \right)}{M(9 M-4 r_{\Sigma^e})}}\right]-\left(\int{\frac{e^{\lambda/2}}{r}dr}\right)_{\Sigma^e}.
\end{equation}

It is worth noticing the difference between (\ref{fint}) and the corresponding expression in the spherically symmetric case (equation (40) in \cite{cf}).

Obviously the conformally flat condition only determines one generating function, therefore in order to find a specific model we have to impose an additional restriction. As an example we shall consider the extreme case $P_r=0$. This solution represents the hyperbolically symmetric analogue of the model I found in \cite{cf} for the spherically symmetric case.

\noindent Thus assuming $P_r=0$, we find from  (\ref{pr})
\begin{equation}\label{nu1n}
  \nu^\prime=-\frac{(1+e^\lambda)}{r}.
\end{equation}

\noindent Replacing  (\ref{nu1n}) in (\ref{PEE}) and using $\mathcal{E}=0$ the following relationship may be easily obtained

\begin{equation}\label{nu2}
  8(1+e^\lambda)+(1+e^\lambda)^2+3r\lambda^\prime-r\lambda^\prime e^\lambda=0,
\end{equation}
or
\begin{equation}\label{nuz}
g(9g-4)-g^\prime r (3g-2)=0,
\end{equation}
where $e^{-\lambda}=2g-1$.
\noindent The integration  of  (\ref{nuz}) produces
\begin{equation}\label{nuzf}
  C_1r^6=\frac{4g^3 }{9g-4},
\end{equation}

\noindent where $C_1$ is a constant of integration.

\noindent Next, combining  (\ref{1}) and (\ref{nu1n}) we obtain
\begin{equation}\label{nug}
e^{\nu}=\frac{\alpha r^2(2g-1)}{g(9g-4)}.
\end{equation}

\noindent For the physical variables we get

\begin{eqnarray}
\vert \mu \vert  &=& \frac{3g}{2\pi r^2}\frac{(2g-1)}{(3g-2)}, \\
  P_\bot &=& \frac{3}{4\pi r^2}\frac{g^2}{(3g-2)}.
\end{eqnarray}
 
From the above it follows that  $g>2/3$ ensuring that $e^{\nu}$ is positive, and defining the minimum value of $r$ for the fluid distribution. The model may be completed by assuming an empty vacuole with boundary surface $r=r_{min}.$ Since  $P_r$ is zero, then both the a.g.m.  and the p.g.m.d. are negative. It should be noticed that while the matching conditions may be satisfied on $\Sigma^e$, the mass function would be discontinuous across $\Sigma^i$. Thus a thin shell appears at the boundary surface $r=r_{min}.$ 

The generating functions for this model are easily found to be

\noindent
\begin{equation}\label{fugez1}
  z=\frac{g-1}{r(2g-1)},
\end{equation}
\noindent and
\begin{equation}\label{fugeP1}
  \Pi(r) = -\frac{3}{4\pi r^2}\frac{g^2}{(3g-2)}.
\end{equation}

\subsection{Two anisotropic solutions from given energy density profiles}
We shall next find two solutions,  by extending  to the hyperbolically symmetric case the procedure developed in \cite{cos} which allows to find an anisotropic solution from any known isotropic one, in the spherically symmetric case.

The basic ansatz of the method is based on a specific form of the anisotropy, more specifically it is assumed  that
\begin{equation}
P_\bot-P_r=C(-\vert \mu \vert +P_r)\frac{\nu^\prime}{2}r,
\label{cos1}
\end{equation}
where $C$ is a constant measuring the anisotropy of the pressure.

Then, using (\ref{cos1}) in (\ref{CPx}) we obtain

\begin{equation}
P^\prime_r+(-\vert \mu \vert +P_r)\frac{\nu^\prime}{2}h=0,
\label{cos3}
\end{equation}
with $h\equiv 1-2C$.

Obviously $h=1$ corresponds to the isotropic pressure case. Then assuming the energy density distribution of a given isotropic solution we may find the corresponding anisotropic model satisfying (\ref{cos1}).

\subsubsection{The incompressible fluid}

Let us first  consider a fluid distribution with constant energy density ($\mu=constant$). If the fluid has isotropic pressure then the solution is unique and would be the hyperbolically  symmetric version of  the interior Schwarzschild solution,  such a solution has been found in \cite{mimII}. However if the pressure is anisotropic there are an infinite number of possible solutions. Here we find a solution which would be the generalization of the Bowers--Liang solution \cite{bl}, for the hyperbolically symmetric fluid.

Thus, assuming $\mu=constant$ we may integrate (\ref{cos3}) obtaining

\begin{equation}
P_r-\vert \mu \vert =\beta e^{-\nu h/2},
\label{cos4}
\end{equation}
where $\beta$ is a constant of integration.

From the junction condition $(P_r)_{\Sigma^e}=0$, we obtain for $\beta$
\begin{equation}
\beta =-\vert \mu \vert e^{\nu_{\Sigma^e} h/2},
\label{cos5}
\end{equation}
producing
\begin{equation}
P_r=\vert \mu \vert \left[1-e^{(\nu_{\Sigma^e}-\nu) h/2}\right].
\label{cos6}
\end{equation}

Also, from (\ref{m2}) we have 
\begin{equation}
m(r)=\frac{4\pi}{3}\vert\mu\vert r^3.
\label{mi}
\end{equation}

Combining (\ref{cos6}) with (\ref{pr}) we obtain
\begin{equation}
\nu^\prime\left(\frac{8\pi r^2\vert\mu\vert}{3}-1\right)-\frac{16\pi \vert\mu\vert r}{3}+8\pi \vert\mu\vert re^{(\nu_{\Sigma^e}-\nu) h/2}=0,
\label{bl1}
\end{equation}
where (\ref{m3n}) has been used.

The integration of (\ref{bl1}) produces
\begin{equation}
e^{\nu h}=\frac{1}{4}\left[3\left(\frac{8\pi \vert \mu \vert r_{\Sigma^e}^2}{3}-1 \right)^{h/2}-\left(\frac{8\pi \vert \mu \vert r^2}{3}-1 \right)^{h/2}\right]^2,
\label{bl2}
\end{equation}
where junction conditions (\ref{j1}) have been used.

Combining  the above expression  with (\ref{cos6}) we obtain,
\begin{equation}
P_r=\frac{\vert \mu \vert  \left[\left(\frac{8\pi \vert \mu \vert r_{\Sigma^e}^2}{3}-1 \right)^{h/2}-\left(\frac{8\pi \vert \mu \vert r^2}{3}-1\right)^{h/2}\right]}{3\left(\frac{8 \pi \vert \mu \vert r_{\Sigma^e}^2}{3}-1\right)^{h/2}-\left(\frac{8\pi \vert \mu \vert r^2}{3}-1\right)^{h/2}},
\label{bl5}
\end{equation}
and for the tangential pressure we have
\\
\begin{widetext}
\begin{equation}
P_\bot=\frac{\vert \mu \vert  \left[\left(\frac{8\pi \vert \mu \vert r_{\Sigma^e}^2}{3}-1\right)^{h/2}-\left(\frac{8\pi \vert \mu \vert r^2}{3}-1\right)^{h/2}\right]}{3\left(\frac{8 \pi \vert \mu \vert r_{\Sigma^e}^2}{3}-1\right)^{h/2}-\left(\frac{8\pi \vert \mu \vert r^2}{3}-1\right)^{h/2}}+ \frac{(1-h) 8\pi \vert \mu \vert ^2 r^2 \left(\frac{8\pi \vert \mu \vert r_{\Sigma^e}^2}{3}-1\right)^{h/2} \left(\frac{8\pi \vert \mu \vert r^2}{3}-1\right)^{\frac{h}{2}-1}}{3\left[3\left(\frac{8 \pi \vert \mu \vert r_{\Sigma^e}^2}{3}-1\right)^{h/2}-\left(\frac{8\pi \vert \mu \vert r^2}{3}-1\right)^{h/2}\right]^2}.
\label{bl6}
\end{equation}
\end{widetext}

As mentioned before, the fluid distribution described above cannot fill the whole space, but is restricted by a minimal value of the $r$ coordinate, satisfying $r_{min}>\sqrt{\frac{3}{8\pi \vert \mu \vert}}$. For $0<r<r_{min}$ we may assume an empty cavity surrounding the center. Also, it is a simple matter to check that both, the a.g.m.  and the p.g.m.d.,  are negative. As in the previous case, this solution may be matched on $\Sigma^e$ but not on $\Sigma^i$, due to the discontinuity of the mass function across that hypersurface.

For  $h=1$ we recover  the incompressible isotropic fluid solution found in \cite{mimII}, whereas for  $h\neq1$ we obtain  the solution equivalent to the Bowers--Liang fluid distribution, corresponding to the hyperbolically symmetric case.

\noindent It is not difficult to see that the two generating functions of this solution are
\begin{equation}\label{fugezCos}
  z=\frac{1}{r}\left [\frac{ 3(\frac{8\pi \vert \mu \vert r_{\Sigma^e}^2}{3}-1)^{h/2}- (\frac{8\pi \vert \mu \vert r^2}{3}-1)^{\frac{h}{2}-1}(\frac{16\pi \vert \mu \vert r^2}{3}-1)}{3(\frac{8\pi \vert \mu \vert r_{\Sigma^e}^2}{3}-1)^{h/2}-(\frac{8\pi \vert \mu \vert r^2}{3}-1)^{h/2}}\right ],
\end{equation}
and
\begin{equation}\label{fugePCos}
  \Pi(r) = -\frac{(1-h)8\pi \vert \mu \vert^2r^2\left(\frac{8\pi \vert \mu \vert r_{\Sigma^e} ^2}{3}-1\right)^{h/2}\left(\frac{8\pi \vert \mu \vert r^2}{3}-1\right)^{\frac{h}{2}-1}}{3\left [3(\frac{8\pi \vert \mu \vert r_{\Sigma^e}^2}{3}-1)^{h/2}-(\frac{8\pi \vert \mu \vert r^2}{3}-1)^{h/2}\right ]^2}.
\end{equation}

\subsubsection{Tolman VI type solution}
As a second application of the approach  sketched above we shall now find a solution inspired in the well known Tolman VI model \cite{tol}.  It is worth recalling that in the spherically symmetric case with isotropic pressure, the  equation of state of this model corresponds to the equation of state of a Fermi gas in the limit of very large energy density.

Thus, following the approach described  above, let us assume
\begin{equation}
\mu=\frac{K}{r^2};\quad \Rightarrow m=-4\pi K r
\label{tol1}
\end{equation}
where $K$ is a negative constant.

Then (\ref{cos3}) for this case may be written as
\begin{equation}
P^\prime_r=\frac{\alpha P_r}{r}+\beta r P^2_r+\frac{\gamma}{r^3},
\label{tol2}
\end{equation}
where the constants $\alpha, \beta, \gamma$  are defined by
\begin{equation}
\alpha\equiv\frac{8\pi K h}{8\pi K+1};\quad \beta\equiv \frac{4\pi h}{8\pi K+1};\quad \gamma\equiv \frac{4\pi h K^2}{8\pi K+1}.
\label{tol3}
\end{equation}

Equation (\ref{tol2}) may be integrated, producing
\begin{equation}
P_r=\frac{(2+\alpha+\epsilon)(2+\alpha-\epsilon)(r^\epsilon_{\Sigma^e}-r^\epsilon)}{2\beta r^2\left[r^\epsilon(2+\alpha-\epsilon)-r^\epsilon_{\Sigma^e} (2+\alpha+\epsilon)\right]},
\label{tol4}
\end{equation}
\\
where  $\epsilon \equiv \sqrt{4+4\alpha+\alpha^2-4\gamma \beta}$, and the boundary condition $P_r(r_{\Sigma^e})=0$ has been used. Thus the matching on $\Sigma^e$ is assured, while the discontinuity of the mass function (\ref{tol1}) across $\Sigma^i$ produces a thin shell on the inner boundary surface.

\noindent For the metric functions and the tangential pressure the corresponding expressions are

\begin{equation}\label{lVI}
  e^{-\lambda} =-(8\pi K+1),
\end{equation}

\begin{equation}\label{nuVI}
  e^\nu=\frac{(\frac{2M}{r_{\Sigma^e}}-1)}{(2\epsilon)^{2n}r_{\Sigma^e} ^{n(2+\epsilon)}}\left\{ r^{n(2-\epsilon)}[r^\epsilon (2+\alpha-\epsilon)-r^\epsilon _{\Sigma^e} (2+\alpha+\epsilon)]^{2n} \right \},
\end{equation}
 with $n=\frac{4\pi}{\beta (8\pi K+1)}$,  and

\begin{widetext}
\begin{eqnarray}
  P_\bot  = \frac{(2+\alpha+\epsilon)(2+\alpha-\epsilon)(r^\epsilon_{\Sigma^e}-r^\epsilon)}{2\beta r^2\left[r^\epsilon(2+\alpha-\epsilon)-r^\epsilon_{\Sigma^e} (2+\alpha+\epsilon)\right]}
   + \frac{2\pi(h-1)}{(8\pi K+1)}\left \{ \frac{r^\epsilon _{\Sigma^e} (2+\alpha+\epsilon)(2-\epsilon)-r^\epsilon (2+\alpha-\epsilon)(2+\epsilon)}{2\beta r \left[r^\epsilon (2+\alpha-\epsilon)-r^\epsilon _{\Sigma^e} (2+\alpha+\epsilon)\right]}\right \}^2.
\end{eqnarray}
\end{widetext}

As it is obvious this fluid distribution is singular at the center,  and therefore  the central region should be excluded by elementary physical reasons.  Furthermore from (\ref{lVI}) it follows that $8\pi\vert K \vert>1$, implying $\alpha>0, \beta<0, \gamma<0, n>0$.

 Finally, for the generating functions of this model we obtain:
\begin{widetext}
\begin{equation}\label{zVI}
  z=\frac{r^\epsilon (2n+n\epsilon+2)(2+\alpha-\epsilon)-r^\epsilon _{\Sigma^e}(2+\alpha+\epsilon)(2n-n\epsilon+2)}{2r[r^\epsilon (2+\alpha-\epsilon)-r^\epsilon _{\Sigma^e} (2+\alpha+\epsilon)]},
\end{equation}
\noindent and
\begin{equation}\label{PiVI}
  \Pi(r)=-\frac{2\pi(h-1)}{(8\pi K+1)}\left \{ \frac{r^\epsilon _{\Sigma^e} (2+\alpha+\epsilon)(2-\epsilon)-r^\epsilon (2+\alpha-\epsilon)(2+\epsilon)}{2\beta r [r^\epsilon (2+\alpha-\epsilon)-r^\epsilon _{\Sigma^e}(2+\alpha+\epsilon)]}  \right \}^2.
\end{equation}
\end{widetext}

\subsection{A model with vanishing complexity factor}
As mentioned before, the scalar $Y_{TF}$ has been shown to be a suitable measure of the complexity of the fluid distribution (see the discussion on this issue in \cite{com}), therefore it would be interesting to find a model (besides the  homogeneous and isotropic solution) satisfying the condition of vanishing complexity ($Y_{TF}=0$). Since there are an infinite number of such solutions, we have to impose an additional restriction in order to obtain a specific model. Here we shall assume (besides the vanishing complexity factor), the condition $P_r=0$.

\noindent Thus, assuming $P_r=0$, we obtain from  (\ref{pr})
\begin{equation}\label{nu1}
  \nu^\prime=-\frac{2g}{(2g-1)r},
\end{equation}
\noindent where $g$ is defined by
\begin{equation}\label{g2}
  e^{-\lambda}=2g-1.
\end{equation}

\noindent Next, imposing  $Y_{TF}=0$ in  (\ref{emtebis}) it follows that

\begin{equation}\label{ytf0}
  m_T=(m_T)_{\Sigma^e} \frac{r^3}{r^3_\Sigma}.
\end{equation}

\noindent The combination of (\ref{masaT1}),(\ref{nu1}),(\ref{g2}) and  (\ref{ytf0})  produces

\begin{equation}\label{nuytf}
  e^\nu=\frac{4(m^2_T)_{\Sigma^e} r^4}{r^6_\Sigma (cosh\pi-1)^2}\frac{(2g-1)}{g^2}.
\end{equation}

On the other hand the condition $Y_{TF}=0$  may be written as
\begin{equation}
g^\prime r (1-g)+g(5g-2)=0,
\label{sol1}
\end{equation}

\noindent whose solution reads
\begin{equation}
C_2r^{10}=\frac{g^5}{(5g-2)^3}\label{g3},
\end{equation}
where $C_2$ is a constant of integration.

\noindent Then for the physical variables we obtain

\begin{eqnarray}
 \vert \mu \vert  &=& \frac{3}{4\pi r^2}\frac{g(2g-1)}{(g-1)}, \\
  P_\bot &=& \frac{3}{8\pi r^2}\frac{g^2}{(g-1)}.
\end{eqnarray}

In this case,  the fluid distribution  is restricted by a minimal value of the $r$ coordinate, satisfying $g(r_{min})>1$. The specific value of $r_{min}$ is obtained from (\ref{g3}). For $0<r<r_{min}$ we may assume, as in precedent models,  an empty cavity surrounding the center. Also as in precedent models, the discontinuity of the mass function across $\Sigma^i$ implies that a thin shell appears on it.  Finally, since the radial pressure is assumed to be zero, both the a.g.m.  and the p.g.m.d. are negative.

The generating functions for this model are easily found to be
\begin{equation}\label{fugez2}
  z=\frac{g-1}{r(2g-1)},
\end{equation}
 and
 \begin{equation}
  \Pi(r) =-\frac{3}{8\pi r^2}\frac{g^2}{(g-1)}.
\end{equation}

\subsection{The stiff equation of state}
Finally we shall consider a couple of  solutions satisfying the so called stiff equation of state, which   as far as we know was proposed for the first time by Zeldovich \cite{zh}, and is believed to be suitable to describe ultradense matter (in particular for neutral vector mesons $\omega^0$ and $\phi^0$). In its original form it assumes that energy density equals pressure (in relativistic units). In our case we shall assume
\begin{equation}
\vert \mu \vert=P_r.
\label{s1}
\end{equation}
then (\ref{m4}) becomes
\begin{eqnarray}
  P_r^\prime+\frac{2}{r} \Pi&=& 0\label{m4z}.
\end{eqnarray}

To obtain specific solutions, additional information is required. Here, as examples, we shall consider two particular cases.
\subsubsection{$P_\bot=0$.}
Let us first assume  that  tangential pressure vanishes. Then (\ref{m4z}) can be easily integrated, producing
\begin{equation}
P_r=\frac{K}{r^2}\Rightarrow \vert \mu \vert=\frac{K}{r^2},
\label{5z}
\end{equation}
where $K$ is a positive constant of integration.

The above equation, together  with (\ref{m3n}), (\ref{m2}) and (\ref{m3}) produces
\begin{equation}
m=4\pi K r,\qquad e^{-\lambda}=8 \pi K-1,\qquad \nu=constant.
\label{z6}
\end{equation}

In this model, both, the a.g.m.  and the p.g.m.d. vanish.

Obviously there are not vanishing pressure surfaces for this solution, and the corresponding generating functions are
\begin{equation}
\Pi=\frac{K}{r^2},\qquad z=\frac{1}{r}.
\label{z7}
\end{equation}
\subsubsection{$Y_{TF}=0$.}

Let us next consider the simplest stiff fluid model (i.e. the one satisfying, besides (\ref{s1}), the  vanishing complexity factor condition).

Then, using this latter condition in (\ref{defYTFbis}) and feeding back the resulting expression into (\ref{m4z}) one obtains

\begin{equation}
P^{\prime \prime}_r+\frac{3}{r}P^\prime_r=0,
\label{z8}
\end{equation}
whose solution reads
\begin{equation}
P_r=\frac{b}{r^2}-a
\label{z9}
\end{equation}
where $a$ and $b$ are two positive constants of integration.

Then from (\ref{m3n}) and (\ref{m2}) it follows at once
\begin{equation}
m=4\pi r\left(b-\frac{ar^2}{3}\right),
\label{z10}
\end{equation}
from which we easily obtain $\lambda$. Finally, feeding back these expressions into (\ref{m3}) we may obtain $\nu$.

Assuming the fluid distribution to be bounded from the exterior by the surface $\Sigma^e$ described by $r=r_{\Sigma^e}=constant$, then we may write
\begin{equation}
P_r=b\left(\frac{1}{r^2}-\frac{1}{{r^2_{\Sigma^e}}}\right),
\label{z11}
\end{equation}

and
\begin{equation}
m= \frac{4\pi b r}{3 r^2_{\Sigma^e}}\left(3 r^2_{\Sigma^e }-r^2\right).
\label{z12}
\end{equation}
Thus, while matching conditions are satisfied on $\Sigma^e$, they are not on $\Sigma^i$.

From the above expressions it follows at once that
\begin{equation}
4\pi r^3P_r-m=-\frac{8\pi br^3}{3 r^2_{\Sigma^e}}.
\label{z13}
\end{equation}

Finally, for the tangential pressure we obtain
\begin{equation}
P_\bot=-\frac{b}{r^2_{\Sigma^e}}
\label{spt}
\end{equation}
Thus,  for this model the p.g.m.d vanishes, whereas,  unlike the previous case, the   a.g.m. $(m_T)$ does not, and   is negative.

\section{Conclusions}
Motivated by the physical interest of the black hole picture described in the Introduction, and which assumes that the spacetime inside the horizon is described by (\ref{w3}), we have carried out  a general study on the properties of  static fluid distributions endowed with hyperbolical symmetry, which eventually could serve as the source of  such spacetime. Thus we have found that such  fluid distributions may be anisotropic in the pressure, with only two main stresses unequal and  the energy density is necessarily negative. Furthermore,  the fluid cannot fill the whole space within the horizon, the central region being excluded. This is so, whether the energy density within the fluid distribution is regular or not. This last result implies that the central region should  consist in a vacuum cavity, or should be described by a different type of source. On the other hand the fact that the fluid distribution cannot attain the center concurs with the result obtained in \cite{2} indicating that no test particle with finite energy can reach the center. 

The violation of the weak energy condition ($\mu<0$), which in turn  implies that the Tolman mass is negative (if $4\pi P_{r} r^3<m$),  requires some discussion.
 
 Let us start by mentioning that in spite of the fact that from classical physics considerations    we expect the energy density to be positive, negative energy densities are often invoked in extremes cosmological and astrophysical scenarios, usually in relation with possible quantum effects, of the kind we could expect within the horizon (see \cite{we1, we2, we3, cqg, pav} and references therein). 
 
 Besides, it is worth recalling that at purely classical level, it has been shown that any spherically symmetric distribution of charged fluid (independently of its equation of state) whose total mass-radius and charge correspond to the observed values of the electron, must have negative energy distribution (at least for some values of the radial coordinate) \cite{bc, cr, hv}.
The possible origin of this intriguing result may be found in a remark by Papapetrou about the finiteness of the total mass of the Reissner-Nordstrom solution  \cite{pap}. Indeed, since the electrostatic energy of a point charge is infinite, the only way to produce a finite total mass is the presence of an infinite amount of negative energy at the centre of symmetry.
Without entering into a detailed analysis of this issue, which is beyond the scope of this manuscript, we speculate that the violation of the weak energy condition might be related to quantum vacuum of the gravitational field.

Next, we recall that in \cite{2} it has been obtained that any test particle within the horizon, for the metric (\ref{w3}), would experience a repulsive force.  In the case of a fluid distribution, this  repulsive nature of the gravitational interaction  was already brought out in Section II as due to the fact that  the a.g.m. (if $4\pi P_{r} r^3<m$) is negative. On the other hand, as we have already mentioned, we expect  the p.g.m.d.  to be negative, or at most zero, which   according to the equivalence principle (stating that the inertial mass equals the passive gravitational mass) would imply that the inertial mass is negative too, therefore a negative pressure gradient even if directed outwardly, would push any fluid element inwardly.  In other words, the forces acting on any fluid element look as if they have switched their role, with respect the  positive energy density case.

Finally, we have developed a general formalism to express any static hyperbolically symmetric fluid solution in terms of two generating functions. Some explicit solutions have been found and their physical variables have been exhibited.  In all of them it appears clearly  that the central region cannot be filled with the fluid distribution. Further study on the physical properties of these solutions, although out of the scope of this work,  would be necessary.

\begin{acknowledgments}
 This work was partially supported by Ministerio de
Ciencia, Innovacion y Universidades. Grant number: PGC2018--096038--B--I00.
\end{acknowledgments}
 
\appendix
\section{Some basic formulae}
In what follows we shall deploy some formulae used  in  our discussion.
\subsection{Christoffel symbols}
\begin{eqnarray}
 \Gamma^0_{10} &=&\frac{\nu^\prime}{2},\quad \Gamma^1_{00}=\frac{\nu^\prime}{2}e^{(\nu-\lambda)},\quad \Gamma^1_{11}=\frac{\lambda^\prime}{2}, \nonumber\\
  \Gamma^1_{22}&=&-re^{-\lambda},\quad \Gamma ^1_{33}  = -re^{-\lambda}\sinh^2\theta,  \\
  \Gamma^2_{12}  &=& \frac{1}{r},\quad \Gamma^2_{33}=-\frac{1}{2}\sinh 2\theta,\nonumber \\
   \Gamma^3_{32}&=&\coth\theta,\quad \Gamma^3_{13}=\frac{1}{r}.\nonumber
\end{eqnarray}

\subsection{Ricci tensor and curvature scalar} 

\begin{eqnarray}
R^0_0 &=& \frac{e^{-\lambda}}{4r}\left[\nu^\prime(4-r\lambda^\prime+r\nu^\prime)+2r\nu^{\prime\prime}\right],\nonumber \\
R^1_1 &= &\frac{e^{-\lambda}}{4r}\left[-\lambda^\prime(4+r\nu^\prime)+r(\nu^\prime)^2+2r\nu^{\prime\prime}\right],\\
  R^2_ 2= R^3_3&=&\frac{e^{-\lambda}}{2r^2}(2+2e^{\lambda}-r\lambda^\prime+r\nu^\prime),\nonumber
 \end{eqnarray}
 \begin{equation}
  {\cal R}=\frac{e^{-\lambda}}{2r^2}\left[4(1+e^\lambda)+ 4r\nu^\prime-4\lambda^\prime r +r^2(\nu^\prime)^2-\lambda^\prime \nu^\prime r^2 +2r^2\nu^{\prime\prime}\right].
\end{equation}

\end{document}